\documentclass[preprint,showpacs,preprintnumbers,amsmath,amssymb]{revtex4}
\usepackage[dvips]{graphics}

\begin{document}
\draft
\title {Finite-dimensional states and entanglement generation for a nonlinear coupler.}
\author {A. Kowalewska-Kud{\l}aszyk}
\email{annakow@amu.edu.pl}
\affiliation{Nonlinear Optics Division, Institute of Physics, Adam Mickiewicz
 University, Umultowska 85, 61-614 Pozna\'n, Poland}
\date{\today}
\author {W. Leo\'nski}
\email{wleonski@amu.edu.pl}
\affiliation{Nonlinear Optics Division, Institute of Physics, Adam Mickiewicz
 University, Umultowska 85, 61-614 Pozna\'n, Poland}
\date{\today}

\begin{abstract}
We discuss a system comprising two nonlinear (Kerr-like) oscillators coupled mutually by a nonlinear interaction. 
The system is excited by an external coherent field that is resonant
to the frequency of one of the oscillators. 
We show that the coupler evolution can be closed within a finite set of $n$-photon states, analogously 
as in the \textit{nonlinear quantum scissors} model. Moreover, for
this type of evolution our system can be treated as a  
\textit{Bell-like states} generator. Thanks to the nonlinear nature of both: oscillators and their internal coupling, 
these states can be generated even if the system exhibits its energy dissipating nature, contrary to systems with linear couplings.
\end{abstract}


\maketitle
\section{Introduction}
In recent years the interest in the possibility of
generation and manipulation of non-classical states has been
growing. It should be stressed that such a possibility is very important 
for the implementation of models in the quantum information theory. 
 In fact, of special interest are the systems, whose dynamics can be closed within a
finite set of $n$-photon states, due to the potential possibilities of
implementations of the models discussed to the quantum 
computational systems. The systems leading to the finite-dimensional
state generation are referred to as linear
\cite{PPB98,BP99} or non-linear \cite{L97, LDT97} {\em quantum scissors}.
Also, nonlinear couplers (NC), {\it i.e.} the systems involving two Kerr nonlinear oscillators with linear interactions
between them, coupled with an external field have been considered in
that context \cite{LM04}. The idea of the NC was introduced
by Jensen \cite{J82} and Maier \cite{Mai82} and has been developed and investigated in many aspects (for a comprehensive
review see \cite{PP00}). For instance, NC systems can exhibit such phenomena as: self-modulation,
self-switching, self-trapping \cite{ChB96}, chaotic systems
synchronisation \cite{GS01} and others. Moreover, NC can exhibit
various quantum-optical features such as sub-Poissonian or squeezed light generations \cite{HSBP89, KP97, FKP99, IUW00}.
What is most interesting from our point of view, these models can also
lead to generation of the finite dimensional and 
maximally entangled states. As we shall show in this paper, the
Bell-like states can also be produced by the NC
systems. However, one should realize that the
systems containing the Kerr nonlinearities are very sensitive to the damping processes  \cite{LT94} and hence,  the losses of various
nature are able to destroy generation of the states discussed almost completely. Nevertheless, we shall show that thanks to the
nonlinear nature of the internal coupling involved, the influence of the damping is not so pronounced as for the systems
with linear couplings \cite{ML05}.

\section{The model and solutions}
The model of a nonlinear coupler considered here is built of two nonlinear
oscillators characterized by Kerr nonlinearities $\chi_{a}$ and
$\chi_{b}$, corresponding to the two cavity field-modes, labelled by $a$ and $b$ respectively. These oscillators are located inside a high-$Q$ cavity and are coupled
together by the interaction of the nonlinear
character. Additionally, the field mode corresponding to the one of the Kerr oscillators (mode $a$) is linearly coupled  to an
external classical field. 
Thus, the system discussed can be described by the effective Hamiltonian (in the interaction picture):
\begin{equation}
\hat{H}=\hat{H}_{NL}+\hat{H}_{int}+\hat{H}_{ext}\,\,\,,
\label{heq1}
\end{equation}
where 
\begin{subequations}
\label{leq2}
\begin{eqnarray}
\hat{H}_{NL}&=&\frac{\chi_a}{2}(\hat{a}^\dagger)^2\hat{a}^2+
\frac{\chi_b}{2}(\hat{b}^\dagger)^2\hat{b}^2\,\,\,,\label{eq2a}\\
\hat{H}_{int}&=&\epsilon (\hat{a}^\dagger)^2\hat{b}^2+
\epsilon^* (\hat{b}^\dagger)^2\hat{a}^2\,\,\,,\label{eq2b}\\
\hat{H}_{ext}&=&\alpha \hat{a}^\dagger+{\alpha}^* \hat{a}\,\,\,.\label{eq2c}
\end{eqnarray}
\end{subequations}
 $\hat{a}(\hat{a}^{\dagger})$ and $\hat{b}(\hat{b}^{\dagger})$,
 appearing above, are the usual bosonic
annihilation (creation) operators, corresponding to the  $a$ and
$b$ modes of the oscillators, respectively. The parameter $\epsilon$
describes the strength of the internal coupling between the two
oscillators, whereas $\alpha$ is the strength of the linear coupling
between the external classical field and the field of the mode $a$ inside the cavity. 

Since this section is devoted to the ideal situation, i.e. the model without damping processes,
we shall describe the system evolution in terms of the time-dependent
wave-function (a more realistic situation of a damped coupler will be also
considered and discussed in the fifth section).
It can be expressed in the $n$-photon Fock basis as:
\begin{equation}
\left|\psi (t)\right>=\sum_{n,m=0}^{\infty} c_{n,m}(t)
\left|n\right>_a
\left|m\right>_b\,\,\,,
\label{eq1}
\end{equation}
where $c_{n,m}(t)$ are the complex probability amplitudes of finding
the system in the $n$-photon state for mode $a$ and the $m$-photon
state for mode $b$.

Since our model involves the external pumping by the classical coherent field, the total
energy of the  system considered is not conserved. Therefore, 
one could expect that with increasing number of photons the Fock
states will
be involved in the system dynamics. However, the assumption of the weak
external pumping and constant amplitudes of the couplings enables us to use the method of the {\em non-linear quantum
scissors} extensively discussed in \cite{L97,LM01}. In consequence, we are able to truncate
the wave function (\ref{eq1}) to the wave function describing only the
evolution of the resonant states. The form of $\hat{H}_{NL}$
implies that only $\left|0\right>_a\left|2\right>_b$, $\left|1\right>_a\left|2\right>_b$ and $\left|2\right>_a\left|0\right>_b$ states
should be considered. 
Thus, the wave function takes the form:
\begin{equation}
\label{eq2}
\left|\psi (t)\right>_{cut}= c_{2,0}(t)\left|2\right>_a\left|0\right>_b+
c_{1,2}(t)\left|1\right>_a\left|2\right>_b+
c_{0,2}(t)\left|0\right>_a\left|2\right>_b\,,
\end{equation}
where we have restricted our considerations to a finite set of the states.

Applying the Schr\"odinger equation we are able to write the set of
equations of motion for three probability amplitudes in the closed form (we use units of $\hbar =1$):
\begin{eqnarray}
\label{eq3}
i\frac{d}{dt}c_{2,0}&=&2\epsilon\,c_{0,2}\,\,\,,      \nonumber\\
i\frac{d}{dt}c_{1,2}&=&\alpha\,c_{0,2}\,\,\,,         \nonumber\\
i\frac{d}{dt}c_{0,2}&=&2\epsilon^*\,c_{2,0}+\alpha^*\,c_{1,2}\,\,\,.
\end{eqnarray}
Assuming that for the time $t=0$ we have two photons in mode $a$ and
no photons in mode $b$ inside the cavity {\em i.e.} 
$\left|\psi(t=0)\right>_{cut}=\left|2\right>_{a}\left|0\right>_{b}$, we can
easily solve the set of equations (\ref{eq3}) showing that the solutions have the form:
\begin{eqnarray}
\label{eq4}
c_{2,0}(t)&=&\frac{\alpha^2+4\epsilon^2\cos(\Omega\,t)}{\Omega^2}\,\,\,,\nonumber
  \\
c_{1,2}(t)&=& 2\,\,\frac{\epsilon\,\alpha\left(\cos (\Omega\,t)-1\right)}
{\Omega^2}\,\,\,,    \nonumber\\
c_{0,2}(t)&=&\,-2i\,\frac{\epsilon\,\sin
  (\Omega\,t)}{\left|\Omega^2\right|}\,\,\,,
\end{eqnarray}
where we have introduced the effective frequency $\Omega=(\alpha^2+4\epsilon^2)^{1/2}$.

To validate our analytical results we can also perform numerical calculations to find the probability amplitudes using the
method described in \cite{LM01}. Thus, we construct the unitary evolution operator
$\hat{U}=\exp\left(-i\hat{H}t\right)$ using Hamiltonian (\ref{heq1}) and apply it to the initial
system state $\left|\psi(t=0)\right>$. For the situation discussed here, we obtain the wave function in the following form:
\begin{equation}
\label{eq5}
\left|\psi (t)\right>=\hat{U}(t)\left|2\right>_a\left|0\right>_b\,\,\,.
\end{equation}
What is important for the numerical calculations performed here, we
used the base including the states
corresponding to much larger numbers of photons than for the states involved in $\left|\psi\right>_{cut}$.
We shall show that the analytical formulae (\ref{eq3}) describing the evolution of the system restricted to the three 
states and the numerical results (with extended base) are in good agreement. This will ensure us
that the applied method of the state truncation
({\em non-linear quantum scissors}) gives correct results.
To prove it, we can analyse the fidelity defined by:
\begin{equation}
\label{eq6}
F=||\left<\psi(t)|\psi(t)\right>_{cut}||^{2}\,\,\, ,
\end{equation}
where $\left|\psi(t)\right>_{cut}$
defined by (\ref{eq2}) is the truncated state and $\left|\psi(t)\right>$ is
the state obtained by the numerical calculations for a large two-mode Hilbert
space. Inspection of the time-evolution of $F$ enables us to estimate the quality of state truncation.
The fidelity for the perfect truncation should give one. In particular, Fig.1 shows the time evolution of the quantity
$1-F$. We can see
that for the case discussed here, the deviations of $F$ from the unity are only about $6\times 10^{-4}$, so
the optical state truncation is fully justified.
\begin{figure}[t]
\begin{center}
\vspace*{-0.1cm}
\resizebox{8,5cm}{6cm}
                { \includegraphics{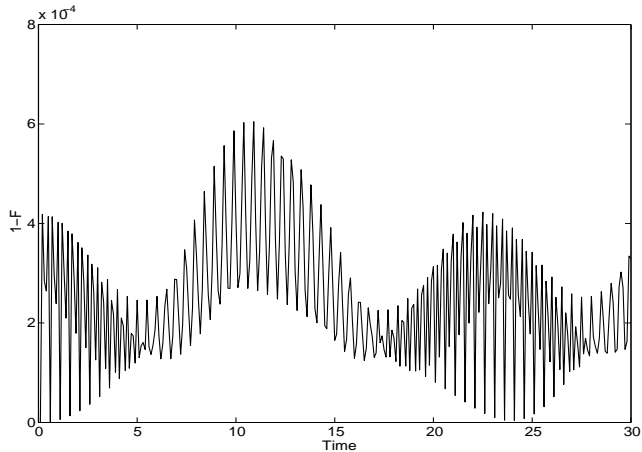}}
\vspace*{-0.5cm} \caption{The time-evolution of the parameter $1-F$.
  The coupling strengths $\alpha =\pi/25$, $\epsilon =\pi /25$, and the nonlinearities
  $\chi_a=\chi_b=25$. Time is scaled in $1/\chi$ units.}
\end{center}
\label{fig1}
\end{figure}
In consequence, our solutions indicate that for mode $a$ the evolution is restricted to states
corresponding to the vacuum, $1$ or $2$ photons, whereas for $b$ the system evolves between
the two states $\left|0\right>_b$ and $\left|2\right>_b$
only. Finally, we can get, to a high accuracy, the state (\ref{eq2}), 
defined in the finite-dimensional Hilbert space, similarly as in \cite{L97}. 
Therefore, the system described here can be referred to as {\em two-mode nonlinear quantum scissors} (TMNQS).
Moreover, from the quantum information theory point of view, our model can be treated as a {\em qutrit}--{\em qubit} system
($\{0,1,2\}\leftrightarrow\{0,2\}$).

\section{Entanglement and the Bell-like states generation}
In the previous section we have shown that the system's dynamics can be closed within a finite set of states,
and hence, can be treated as TMNQS. This section is devoted to another very interesting and important feature, namely the system's
ability to generate the maximally entangled sates (MES).

Since the system evolution can be described in terms of the qutrit-qubit system, one can expect that
for the model discussed here the Bell-like states can be generated.
To investigate this phenomenon we plot the probabilities for the states involved in our model. Thus, 
Fig.2 shows that these probabilities oscillate and some of the plots intersect for the values close to $0.5$.
This indicates that the Bell-like states could be generated in the situation discussed here.  In particular, we observe
this crossing for the probabilities corresponding to the states $\left|2\right>_a\left|0\right>_b$ and $\left|0\right>_a\left|2\right>_b$. Therefore, we shall concentrate
on the states:
\begin{eqnarray}
\label{eq7}
\left|B_1\right>&=&\frac{1}{\sqrt{2}}\left(\left|2\right>_a\left|0\right>_b
+i\,\left|0\right>_a\left|2\right>_b\right)\,\,\,, \nonumber \\
\left|B_2\right>&=&\frac{1}{\sqrt{2}}\left(\left|2\right>_a\left|0\right>_b
-i\,\left|0\right>_a\left|2\right>_b\right)\,\,\,.
\end{eqnarray}
Moreover, we see that the probabilities corresponding to $\left|1\right>_a\left|2\right>_b$ and $\left|0\right>_a\left|2\right>_b$ exhibit the same behaviour
(crossing for the values close to $0.5$). This corresponds to
generation of the states $\left|P_1\right>$ and $\left|P_2\right>$, defined as follows:
\begin{eqnarray}
\label{eq8}
\left|P_1\right>&=&\frac{1}{\sqrt{2}}\left(\left|1\right>_a\left|2\right>_b
+i\,\left|0\right>_a\left|2\right>_b\right)\,\,\,, \nonumber \\
\left|P_2\right>&=&\frac{1}{\sqrt{2}}\left(\left|1\right>_a\left|2\right>_b
-i\,\left|0\right>_a\left|2\right>_b\right)\,\,\,.
\end{eqnarray}
It is seen that these states can be expressed as 
\begin{equation}
\left|P_{1,2}\right>=\frac{1}{\sqrt{2}}\left(\left|1\right>\pm i\left|0\right>\right)_a\otimes\left|2\right>_b\,\,\,,
\label{eq9}    
\end{equation}
where we get the product of two states, defined within various subspaces corresponding to
 modes $a$ and $b$, respectively. In consequence, we cannot expect
 entanglement when the system reaches the states $\left|P_{1,2}\right>$ and 
the product state is generated.

To find whether the system can produce the states we are interested in, we plot (Fig.3) the fidelities corresponding to the states
$\left|B_{1,2}\right>$ and $\left|P_{1,2}\right>$, \textit{i.e.}
$|\left<\psi(t)|B_{1,2}\right>|$ and $|\left<\psi(t)|P_{1,2}\right>|$
respectively. We see, that for the properly chosen moments of time,
these states can be generated to a high accuracy
and some of the maxima in Fig.3a are very close to unity. Moreover, we see that the fidelities corresponding to the Bell-like
states reach other maximal values, which are however, close to $\sim 0.7$. This indicates the existence of entanglement in the system,
however, we get the states that are not $100\%$ entangled. However if we enhance the coupling between the oscillators, the probability of creation of
the state  $\left|1\right>_a\left|2\right>_b$ decreases significantly and as a consequence, the fidelities for formation of the states $\left|B_{1,2}\right>$ are almost equal to unity (Fig.(3b) -- solid line). Moreover, the oscillations between the states become faster and the system tends to be a qubit-qubit one.
\begin{figure}[t]
\centering
\vspace*{-0.1cm}
\resizebox{8,5cm}{6cm}
                { \includegraphics{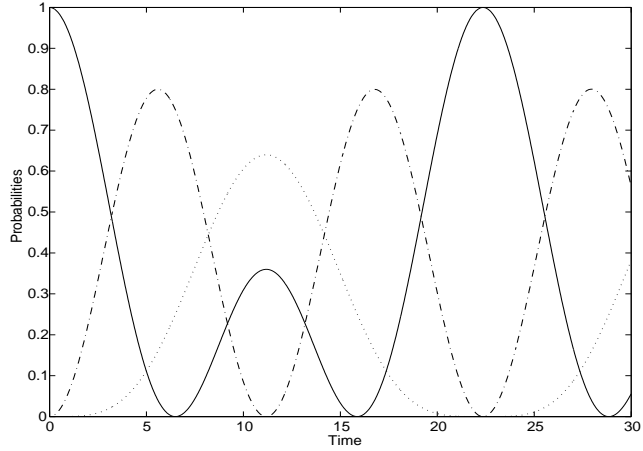}}
\vspace*{-0.5cm}
\caption{The probabilities for the states $\left|2\right>_a\left|0\right>_b$ (solid line), $\left|0\right>_a\left|2\right>_b$ (dashed-dotted line)
and $\left|1\right>_a\left|2\right>_b$ (dotted line). The coupling strengths $\alpha =\pi/25$, $\epsilon =\pi /25$, and the nonlinearities
  $\chi_a=\chi_b=25$. Time is scaled in $1/\chi$ units.}
\label{fig2}
\end{figure}
\begin{figure}[t]
\centering
\vspace*{-0.1cm}
\resizebox{7cm}{6cm}
                { \includegraphics{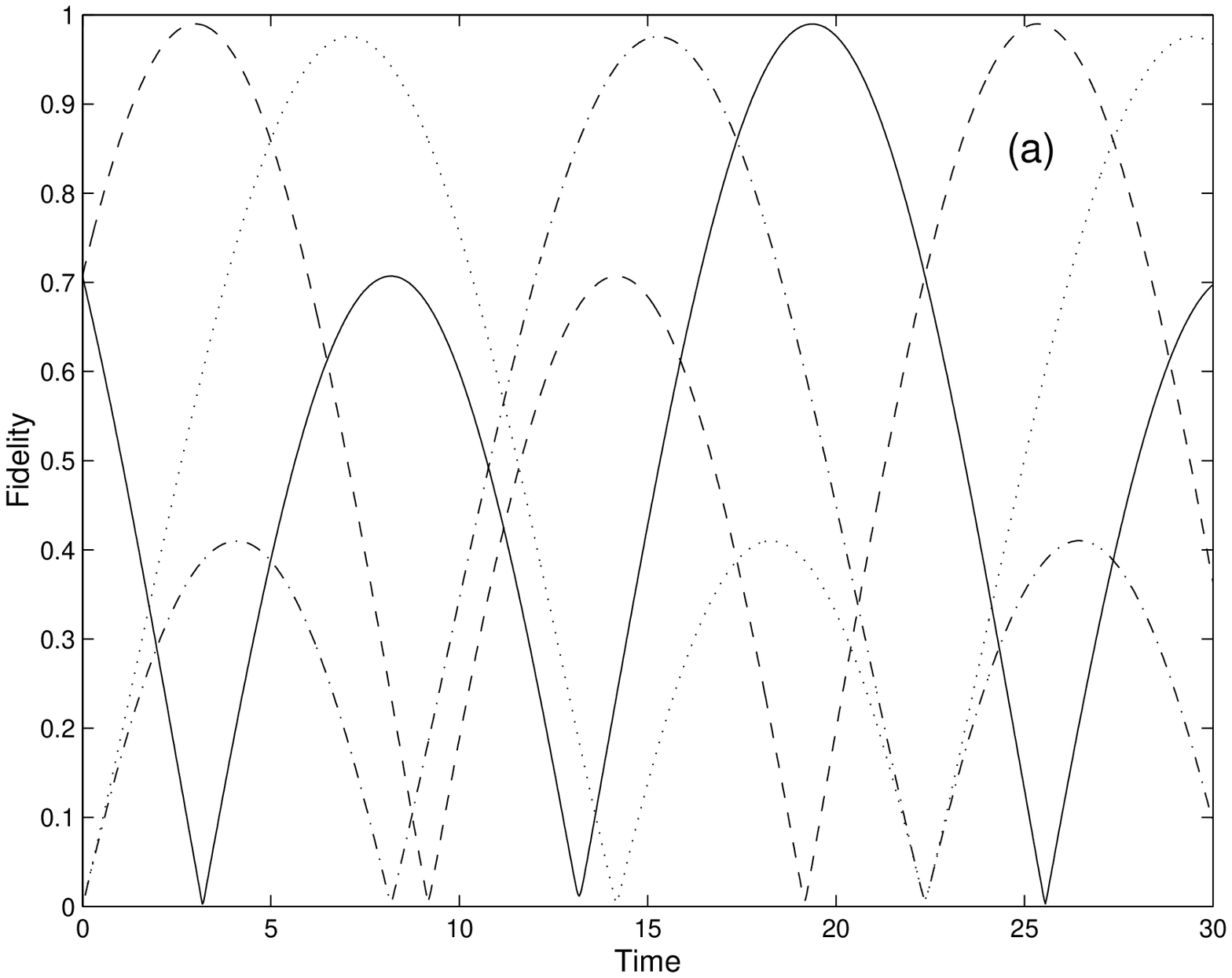}}
\hspace{0.5cm}
\resizebox{7cm}{6cm}
                { \includegraphics{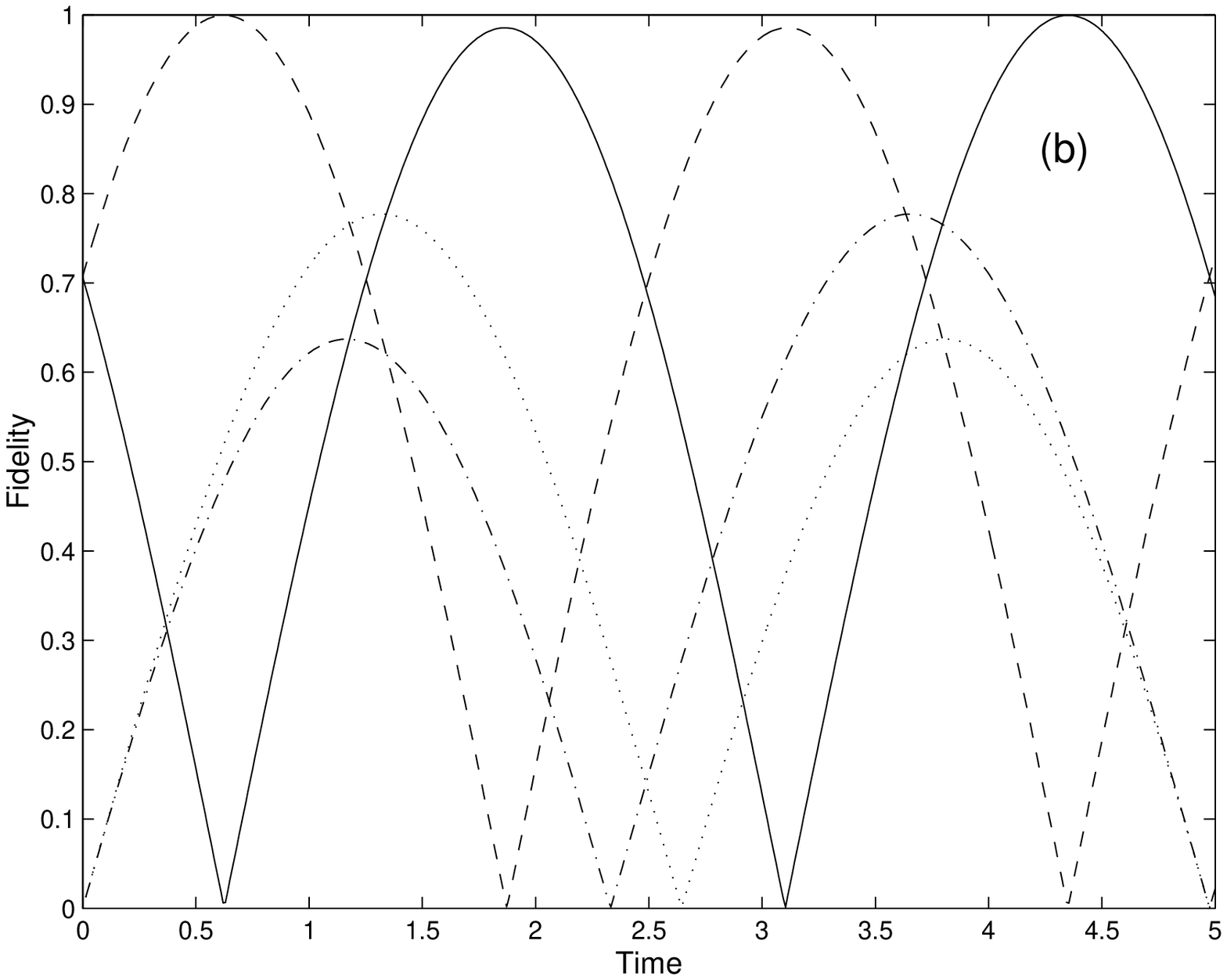}}
\vspace*{-0.5cm}
\caption{The fidelities corresponding to the Bell-like states $\left|B_{1}\right>$ (solid line), $\left|B_{2}\right>$ (dashed line),
and the product states $\left|P_1\right>$ (dotted line), $\left|P_2\right>$ (dashed-dotted line). The parameters at a) are identical to those of 
Fig.2. b) differ in $\epsilon=\pi/5$.}
\label{fig3}
\end{figure}

It is obvious that we could inspect other combinations of the pure states discussed here. For instance, the Bell-like states involving $\left|1\right>_a\left|2\right>_b$ and
$\left|2\right>_a\left|0\right>_b$ could be considered.
After inspection of the results plotted in Fig.2, we see that these
Bell-like states could play some role in the system's evolution. We
observe the crossings of the plots of the appropriate probabilities,
however, their points of intersection correspond to the value of the probability equal to $\sim 0.1$. Therefore, for these moments of time the system is practically in the pure state $\left|0\right>_a\left|2\right>_b$.
Nevertheless, one can see that for some other moments of time these probabilities become close to $1/2$, although they do not intersect. Consequently, one can expect Bell-like states generation again. Fig.4 shows the fidelities corresponding to the following Bell-like states:
\begin{eqnarray}
\left|B_3\right>&=&\frac{1}{\sqrt{2}}\left(\left|2\right>_a\left|0\right>_b
+i\,\left|1\right>_a\left|2\right>_b\right)\,\,\,, \nonumber \\
\left|B_4\right>&=&\frac{1}{\sqrt{2}}\left(\left|2\right>_a\left|0\right>_b
-i\,\left|1\right>_a\left|2\right>_b\right)\,\,\,, \nonumber \\
\left|B_5\right>&=&\frac{1}{\sqrt{2}}\left(\left|2\right>_a\left|0\right>_b
+\,\left|1\right>_a\left|2\right>_b\right)\,\,\,, \nonumber \\
\left|B_6\right>&=&\frac{1}{\sqrt{2}}\left(\left|2\right>_a\left|0\right>_b
-\,\left|1\right>_a\left|2\right>_b\right)\,\,\,.
\label{eq10}
\end{eqnarray}
\begin{figure}[t]
\centering
\vspace*{-0.1cm}
\resizebox{8,5cm}{6cm}
                { \includegraphics{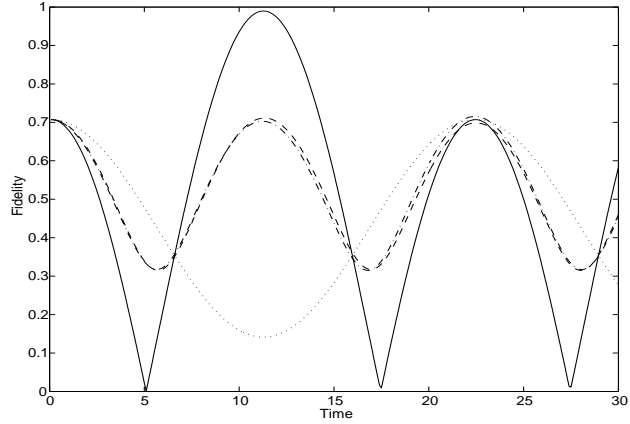}}
\vspace*{-0.5cm}
\caption{The fidelities corresponding to the Bell-like states: $\left|B_{3}\right>$ (dashed-dotted line), $\left|B_{4}\right>$ (dashed line),
$\left|B_{5}\right>$ (solid line), $\left|B_{6}\right>$ (dotted line). All parameters are identical to those of Fig.2.}
\label{fig4}
\end{figure}
We see that for the time $\sim 12$ the fidelity corresponding to the state $\left|B_{5}\right>$ becomes almost equal to unity. Hence, we get the Bell-like state $\left|B_{5}\right>$ slightly perturbed by the pure state $\left|2\right>_a\left|0\right>_b$.

The ability for generation of maximally entangled states can be described in terms
of the von Neumann entropy $\mathit{E}$ of the reduced density matrix $\rho_a=\mathrm{Tr}_b\rho_{ab}$ or
$\rho_b=\mathrm{Tr}_a\rho_{ab}$ constructed on the basis of the full density matrix
$\rho_{ab}=\left|\psi\right>\left<\psi\right|$ corresponding to
the evolution of the whole system. Equivalently, we can use the Shannon entropy $\mathit{H}$ for the squared Schmidt
coefficients $p_i$
\begin{eqnarray}
\label{eq11}
\mathit{E}&=&-\mathrm{Tr}\rho_{a}\log_{2}\rho_{a}=-\mathrm{Tr}\rho_{b}\log_{2}\rho_{b}\nonumber\\
&=&\mathit{H}\{p_i\}=-\sum_{i}p_{i}\log_{2}p_{i}\,\,\, .
\end{eqnarray}
The entropy changes its value and is equal to zero for separable states, whereas for the maximally entangled states
it should be equal to 1 ebit.
As we see from Fig.5, the entropy reaches a limit of 1 ebit for several times for which the Bell-like (or close to them) states are formed.
So, it proves that the NC with nonlinear interactions between field modes can
be treated as a source of maximally entangled states.
\begin{figure}[h]
\label{entropy}
\begin{center}
\vspace*{-0.1cm}
\resizebox{8.5cm}{6cm}
                {\includegraphics{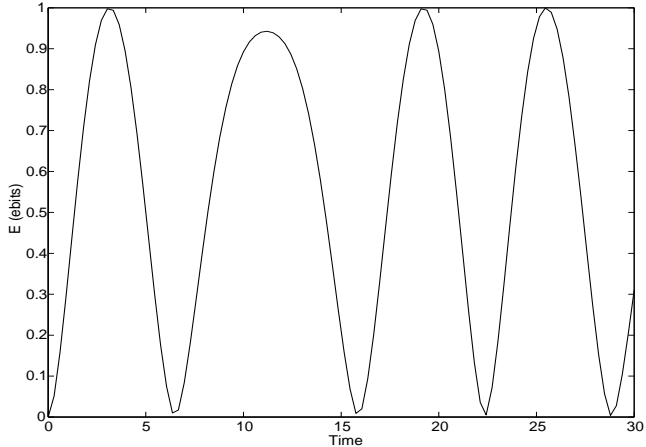}}
\vspace*{-0.5cm} 
\caption{The entropy of entanglement, in ebits, for the
  same parameters as in the previous figures.}
\end{center}
\end{figure}

\section{Evolution of nonlocality}
Both the entanglement of quantum states and the quantum nonlocality are essential and intriguing aspects of quantum mechanics. 
Quantum nonlocality manifests itself by a violation of the Bell-type inequalities \cite{Bell64,CHSH69}. 
It has been demonstrated that states which violate the Bell-type inequalities are the entangled pure states \cite{G91} or a mixture of Bell states. 
There exist also states (known as Werner ones \cite{W89}) which are entangled nevertheless they do not violate any of the Bell inequalities. It is also possible 
to obtain quantum states for which we achieve the nonlocality without any entanglement \cite{HHH99}.

Another problem is the way we can study the quantum nonlocality and its evolution. For this purpose we have usually used various specific parameters obtained numerically. 
For a two-qubit system  we can examine the quantum nonlocality analysing the Bell-CHSH type inequality \cite{CHSH69}
\begin{equation}
|Tr(\rho B_{CHSH})|\leq 2\,\,\, ,
\label{non1}
\end{equation}
where Bell operator $B_{CHSH}$ has a form:
\begin{equation}
B_{CHSH}=\mathbf{a}\cdot\mathbf{\sigma}\otimes(\mathbf{b}+\mathbf{b'})\cdot\mathbf{\sigma}+\mathbf{a'}\cdot\mathbf{\sigma}\otimes(\mathbf{b}-\mathbf{b'})\cdot\mathbf{\sigma}\,\,\, .
\label{non2}
\end{equation}
$\mathbf{a}$, $\mathbf{a'}$, $\mathbf{b}$, $\mathbf{b'}$ are the unit vectors in $\Re^3$ and $\mathbf{\sigma}$ is a vector of Pauli spin matrices 
$\sigma_i$ ($i=1,2,3$). It is known that any state $\rho$ can be expressed as:
\begin{equation}
\rho=\frac{1}{4}\left(I\otimes I+\mathbf{r}\cdot\mathbf{\sigma}\otimes I+I\otimes \mathbf{s}\cdot\mathbf{\sigma}+\sum_{n,m=1}^{3}t_{nm}\sigma_n\otimes\sigma_m\right)\,\,\, ,
\label{non3}
\end{equation}
where $\mathbf{r}$, $\mathbf{s}$ are also the unit vectors in $\Re^3$, scalar product of $\mathbf{r}\cdot\mathbf{\sigma}$ should be understood as 
$\sum_{i=1}^3 r_i\sigma_i$ and the coefficients $t_{nm}=Tr(\rho\sigma_n\otimes\sigma_m)$ form a real matrix $T_\rho$. If we also define the real symmetric 
matrix $U_\rho=T^T_\rho T_\rho$ ($T^T_\rho$ is the transposition of matrix $T_\rho$) we can use for further considerations the Miranowicz measure 
of nonlocality \cite{M04}:
\begin{equation}
B(\rho)\equiv\sqrt{max\{0,M(\rho)-1\}}\,\,\, .
\label{non4}
\end{equation}
$M(\rho)$ as a measure of violation of inequality (\ref{non1}) has been proposed in \cite{HHH95,H96} and takes the following form: $M(\rho)=max_{j<k}\{u_j+u_k\}$, where $u_i$ ($i=1,2,3$)
are the eigenvalues of matrix $U_\rho$. The Bell-CHSH inequality is then violated by the density matrix $\rho$ if $M(\rho)>1$. One can also study other parameters, for instance $n(\rho)=\max \{0,M(\rho)-1\}$, proposed in \cite{JJ03}. The parameter $B(\rho)$ defined in (\ref{non4}) 
varies from zero to unity and is equal to unity for the case of maximal violation of the inequality (\ref{non1}), and its value corresponds to the degree of Bell-CHSH inequality violation.
As our model can be treated as a qutrit-qubit system we shall use \cite{LX05} a local projection operation 
$\Pi_{0,2}\otimes\Pi_{0,2}\equiv(\left|0\right>\left<0\right|
+\left|2\right>\left<2\right|)\otimes(\left|0\right>\left<0\right|
+\left|2\right>\left<2\right|)$ to obtain the density operator for the qubit-qubit system. As a consequence, we get the reduced density matrix $\hat\rho^{(c)}$ of the form:
\begin{equation}
\hat{\rho}^{(c)}=\frac{\Pi_{0,2}\otimes\Pi_{0,2}\hat{\rho}\Pi_{0,2}\otimes\Pi_{0,2}}{Tr\left(\Pi_{0,2}\otimes\Pi_{0,2}\hat{\rho}\Pi_{0,2}\otimes\Pi_{0,2}\right)}\,\,\, .
\label{non5}
\end{equation}
Here, we are able now to analyse the degree of quantum nonlocality using (\ref{non4}) for the two qubit system obtained from two qutrit system by the above projection. As an example, we show the time dependence of the $B(\rho)$ applied for $\hat\rho^{(c)}$ and trace its value for the times where Bell-like states (eq.(\ref{eq7})) are formed.
\begin{figure}[h]
\label{BIV}
\begin{center}
\vspace*{-0.1cm}
\resizebox{7cm}{6cm}
                {\includegraphics{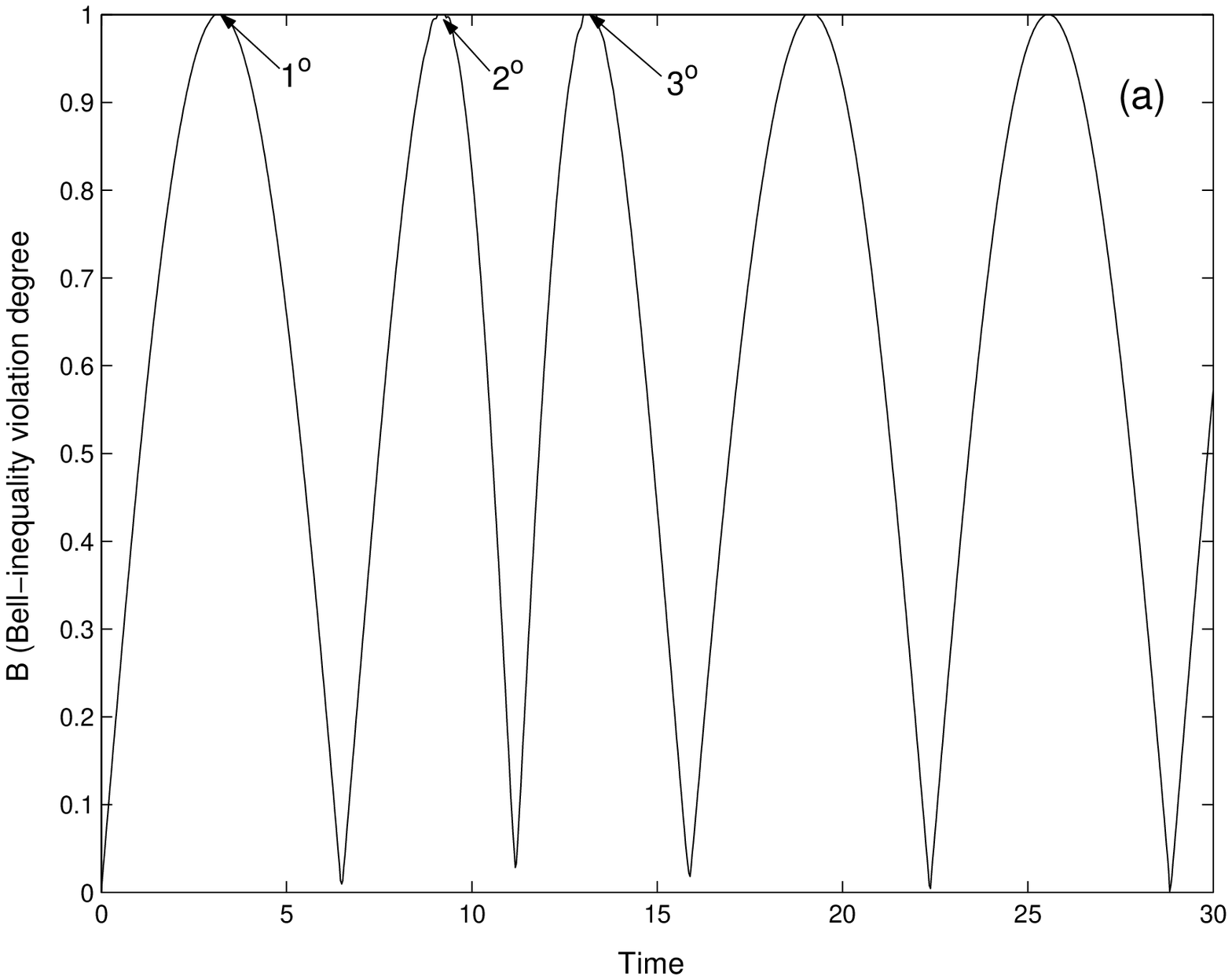}}
\hspace{0.5cm}
\resizebox{7cm}{6cm}
                {\includegraphics{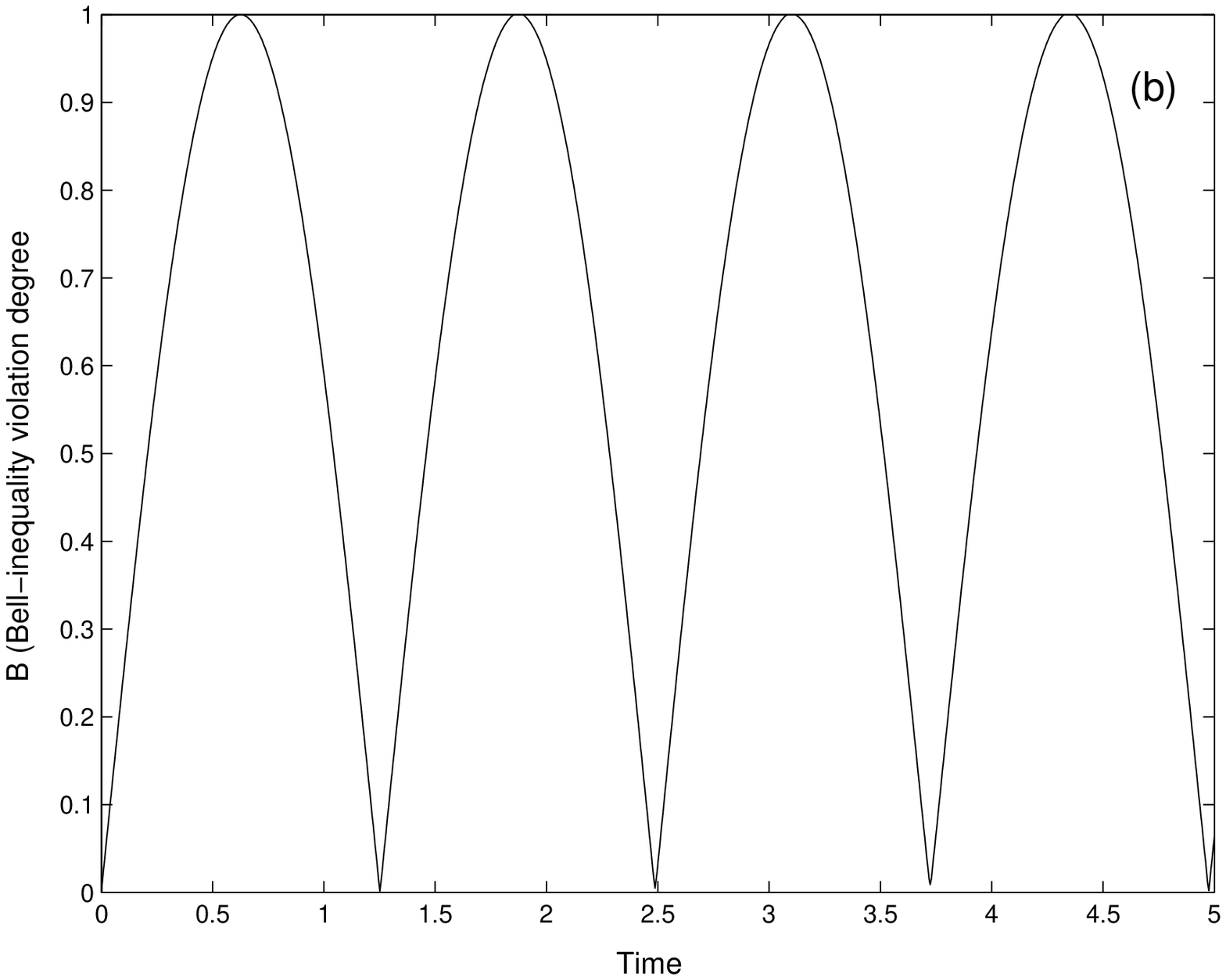}}
\vspace*{-0.5cm} 
\caption{The Bell inequality violation degree corresponding to the formation of Bell-like states $\left|B_{1,2}\right>$ for the coupling strengths $\epsilon =\pi/25$ (a) and $\epsilon =\pi /5$ (b). Time is scaled in $1/\chi$ units. The labels $\{1^0,2^0,3^0\}$ enumerate the maxima -- {\it see the discussion in the text}.}
\end{center}
\end{figure}
Comparing this figure with fig.3a we can easily see that the maximal degree of Bell-CSHS inequality violation appears when one of the Bell-like states
(eq.(\ref{eq7})) is generated with perfect fidelity. The maximal violation also appears when one of the Bell-like states is formated with less fidelity 
(maxima labelled as $2^0$ and $3^0$ at the fig.(6a)). Maximum $2^0$ for example appears when the fidelity of formation of $\left|B_2\right>$ state is zero but at the same time there appears
a state $\left|P_1\right>$ with large fidelity -- however, $\left|P_1\right>$ has also the $\left|0\right>_a\left|2\right>_b$ component. Thus, 
we would say that the influence of the states specified in eq.(\ref{eq8}) give rise to that nonlocality effects. Larger values of the coupling strength result 
in faster oscillations of the parameter $B(\rho)$ considered, similarly as for the fidelities shown in Fig.3b. In consequence, the Bell-CSHS inequality is maximally violated when the Bell-like states $\left|B_{1,2}\right>$
form (Fig.(6b)).

\section{The damping case}
As mentioned earlier, the damping processes are able to modify physical properties of  systems
containing nonlinear oscillators \cite{LT94}. In particular, the ability of finite dimensional states formation
could be considerably reduced. Therefore, we shall concentrate in this section on the influence of
damping processes on the Bell-like states generation,
and we will inspect the effect of the damping process
(caused mostly by the losses at cavity mirrors) on the formation of
maximally entangled states. Therefore, we will have to include into the description of the
system dynamics the appropriate quantities corresponding to the energy
leakage from the cavity.
It is obvious that this problem concerns both $a$ and $b$ modes.
Consequently, we have to define two collapse operators corresponding to the energy losses in these two modes. They are:
\begin{eqnarray}
\label{eq12}
\hat{C}_{1}&=&\sqrt{2\kappa_{a}}\hat{a}\,\,\, ,\nonumber\\
\hat{C}_{2}&=&\sqrt{2\kappa_{b}}\hat{b}\,\,\, .
\end{eqnarray}
We will now turn to the standard methods of the quantum theory of
damping, to find the evolution of the damped system by numerically
solving the master equation for the density matrix.
The generic form of the master equation is
\begin{equation}
\label{eq13}
\frac{d\hat{\rho}}{dt}={\hat{\cal L}}\hat{\rho}\,\,\, ,
\end{equation}
where the Liouvillian $\hat{\cal L}$ is a superoperator and in our case
it takes the form:
\begin{equation}
\label{eq14}
{\hat{\cal
L}}=\frac{1}{i}\left(\hat{H}\hat{\rho}-\hat{\rho}\hat{H}\right)+\sum_{k=1}^{2}\left[\hat{C}_{k}\rho\hat{C}^{\dagger}_{k}-\frac{1}{2}\left(\hat{C}^{\dagger}_{k}\hat{C}_{k}\rho-\rho\hat{C}^{\dagger}_{k}\hat{C}_{k}\right)\right]\,\,\, .
\end{equation}
Once we have found the Liouvillian, we are able to solve the master equation
and get the time evolution of the density matrix elements, which corresponds
to the states involved in the dynamics of the coupler described. The
evolution is found as a series of complex exponentials $\exp({s_{j}t})$,
where each $s_{j}$ is an eigenvalue of $\hat{\cal L}$.

Although it is possible to examine the effect of the damping
processes on the generation of all the entangled states discussed
here, we shall concentrate in this section on the generation of
$\left|B_{1}\right>$ and $\left|B_{2}\right>$ states. We
assume that the processes discussed here perturb the mechanisms of other
Bell-like states generation in the same way. Moreover, we shall compare our results with those concerning NC with a linear internal coupling \cite{ML05}. Therefore, we have assumed that the values of the parameters describing our system are comparable with those discussed in \cite{ML05} -- the values of the parameters discussed there are assumed to be more realistic from the experimental point of view.
\begin{figure}[h]
\label{fig7}
\begin{center}
\vspace*{-0.1cm}
\resizebox{9cm}{5cm}
{\includegraphics{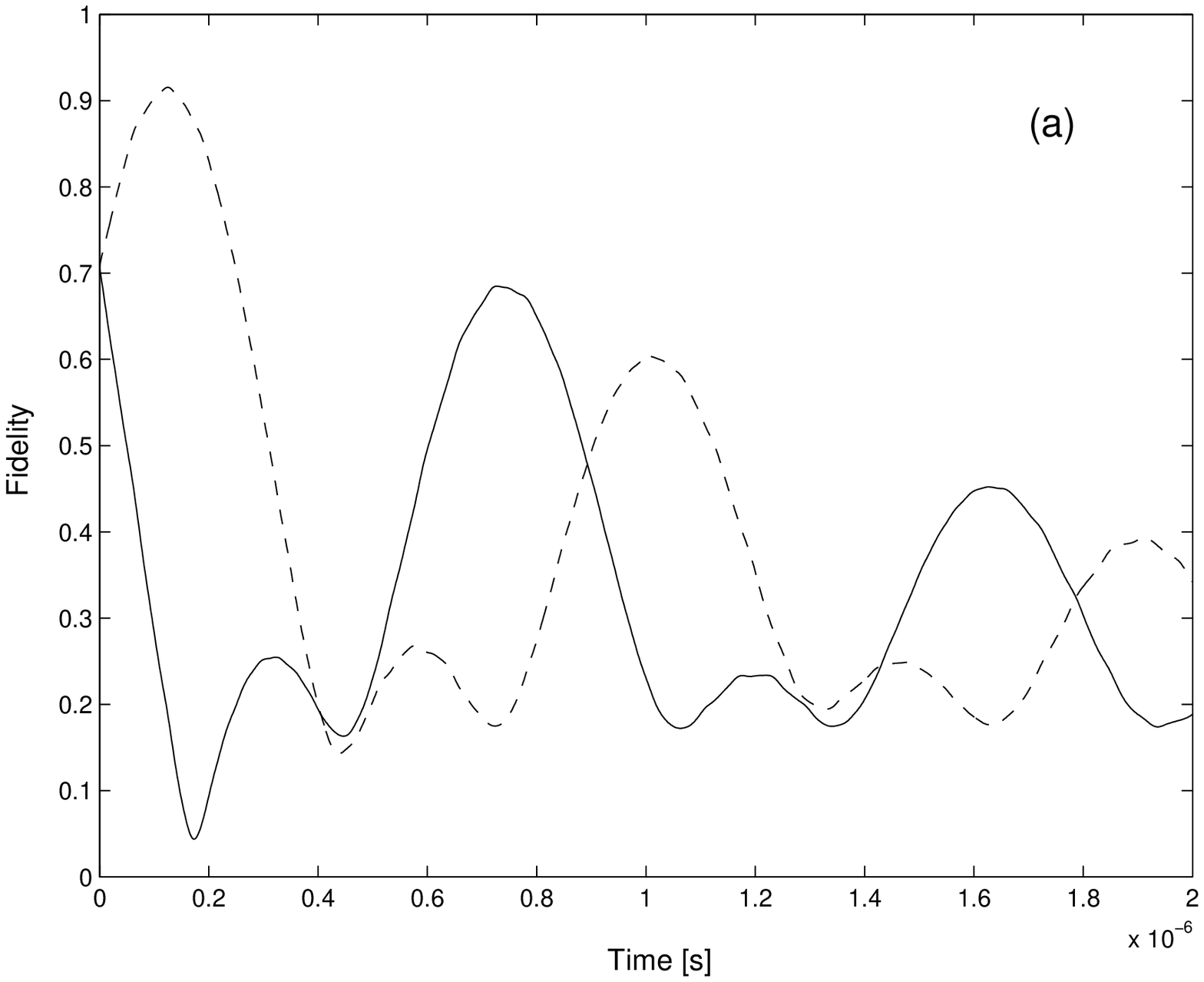}}\\
\resizebox{9cm}{5cm}
{\includegraphics{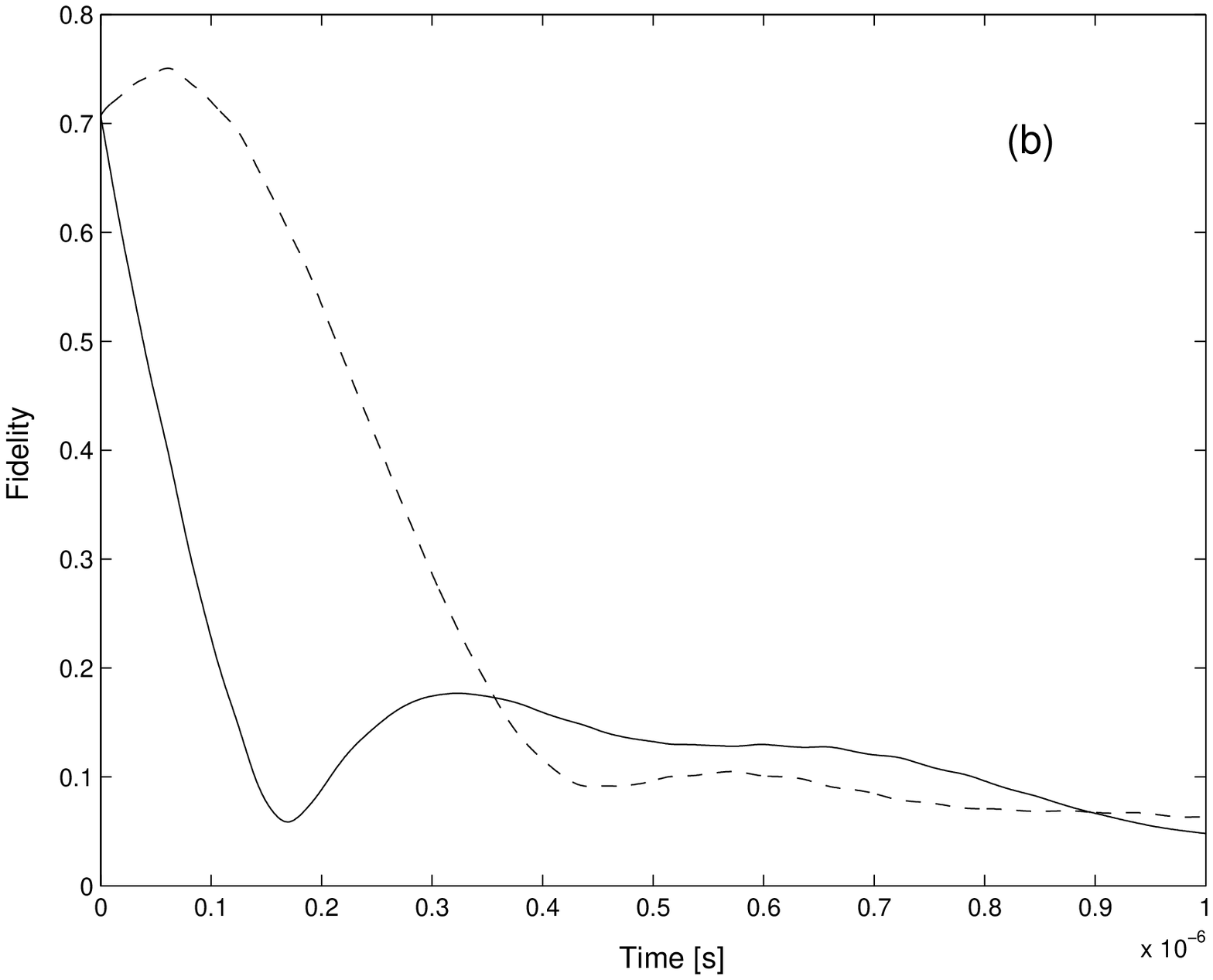}}\\
\resizebox{9cm}{5cm}
{\includegraphics{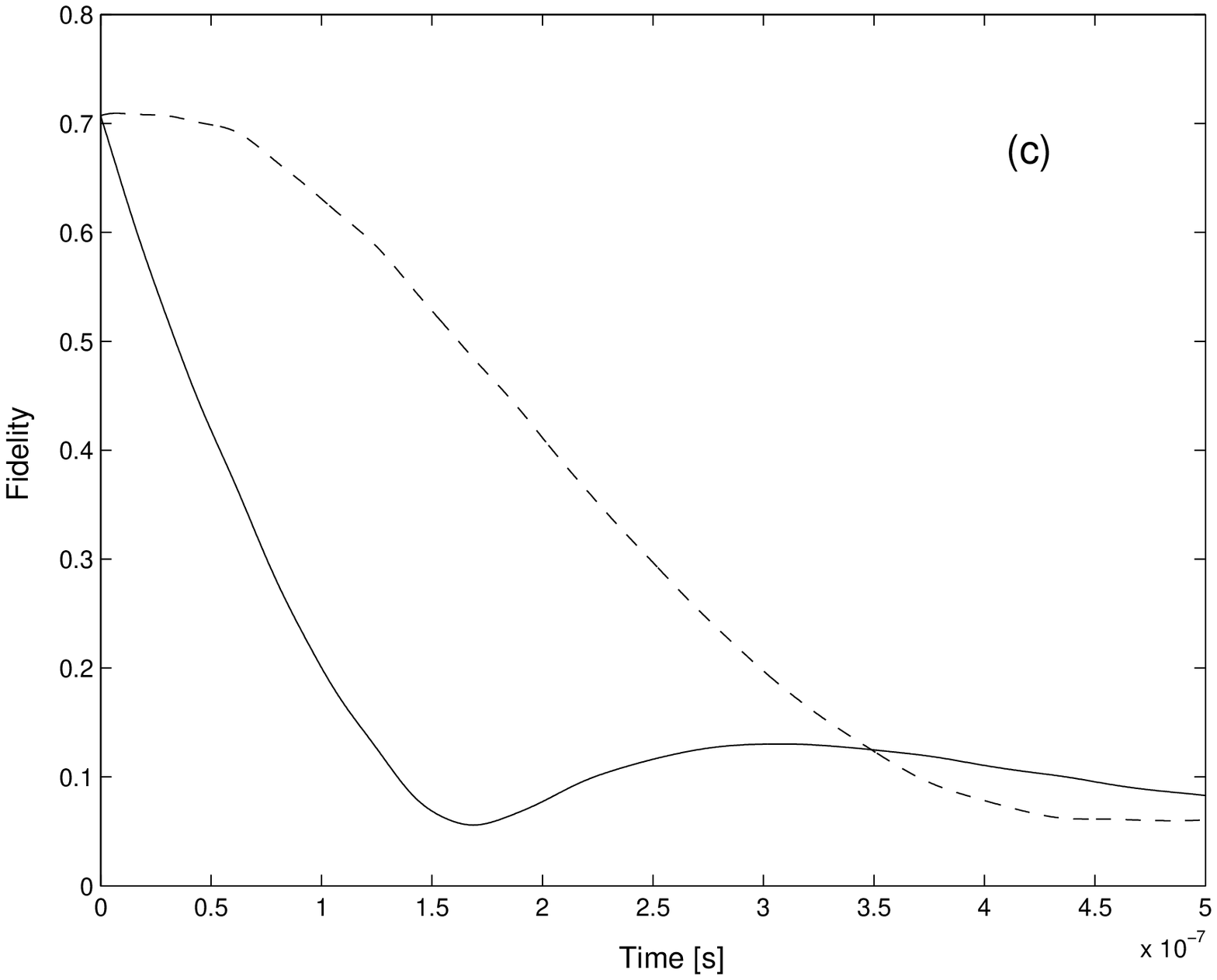}}
\vspace*{-0.5cm} 
\caption{Fidelities corresponding to the Bell-like states $\left|B_1\right>$ (solid lines) and $\left|B_2\right>$ (dashed lines) for various values of the damping constants $\kappa_a=\kappa_b=\kappa$:
$\kappa=\chi/500$ (a),
$\kappa=\chi/75$ (b), and $\kappa=\chi/50$ (c).
The remaining parameters are: $\chi_a=\chi_b=\chi=10^8$rad/sec, $\alpha=\chi/20$, $\epsilon=\alpha/2$.
}
\end{center}
\end{figure}
Since this part of the paper is devoted to the damped case and we use the description of the system based on the density matrix formalism, we use the following definition of the fidelity $F$:
\begin{equation}
F=Tr\sqrt{\rho ^{1/2}\rho_{cut}\rho^{1/2}}\,\,\,,
\label{eq16}
\end{equation}
where $\rho$ and $\rho_{cut}$ are the density matrices corresponding to the states compared.
For the case discussed in the previous section this fidelity becomes identical with that defined in eq.(\ref{eq6}).

Thus, Fig.7 shows the fidelities corresponding to the states
$\left|B_1\right>$ and $\left|B_2\right>$ for different values of the
damping parameters. We see that for a very weak damping case we get
Bell-like state $\left|B_2\right>$ almost perfectly (the same
situation can be observed for the model discussed in
\cite{ML05}). However, if we increase the values of the damping
parameters the effectiveness of the process becomes
considerably reduced. Nevertheless, we see that for $\kappa=\chi/75$ we get
the desired state quite effectively. One should note that for this
case, not only the maximal value of the fidelity corresponding to the
state $\left|B_2\right>$ is important, but also the rate of
simultaneous vanishing of the fidelity for the state
$\left|B_1\right>$ should be considered, due to the fact that the sum
of these states gives the pure state
$\left|2\right>_a\left|0\right>_b$. For the case of $\kappa=\chi/75$
both conditions are sufficiently fulfilled .
The strengths of the damping discussed here are effectively greater
than in \cite{ML05} and consequently, our model seems to be more
suitable for the Bell-like states generation, as it is less sensitive
to the losses in the system. We can explain this fact as a result of
the resonant nature of the nonlinear internal coupling. The coupling
with the external reservoir has a linear character and involves the
one-photon and vacuum states corresponding to two modes $a$ and
$b$, and then the states discussed here
($\{\left|0\right>_a\left|2\right>_b$,\,
$\left|2\right>_a\left|0\right>_b$,\,
$\left|1\right>_a\left|2\right>_b\}$). Consequently, due to the
resonant character of the transitions inside the set of these three
states, the transitions become dominant and couplings to the
states interacting with the external bath can be neglected for the
cases considered here. This result is a consequence of the nature of
our system that is constructed on the \textit{nonlinear scissors}
model base, where the resonances play a crucial role \cite{LDT97}.
Once the damping process causes the decay of the fidelity for 
the Bell-states, one can think about the possibility of applying a 
purification protocol to reduce the influence of the damping. It has been already 
shown \cite{K98,HHH99} that there is no local purification scheme which produces 
a maximally entangled state from a mixed state entangling two finite-dimensional systems. 
Therefore, one should concentrate on the global purification techniques. For these cases 
the model discussed here would be only a part of a greater system, in which the information would be shared between two parties and measured by them. 
Moreover, one should keep in mind that practically all systems that are discussed in that context in literature are two-qubit systems, while the ours 
is a qutrit-qutrit one. 
However, Alber {\it et.al} \cite{ADGJ01} have discussed higher dimensional systems, and this fact gives us some hope for finding the appropriate 
purification protocol that would help improve the maximally entangled state generation mechanism. They have proposed an application of generalized 
$XOR$-gate that can be constructed on the basis of nonlinear Kerr interaction governed by the Hamiltonian $\hat{H}=\chi\hat{n}_1\hat{n}_2$ 
(that differs from those discussed in our paper). 
In the context of the purification protocol presented there, the model discussed here can be treated as promising proposal for further investigation 
of the methods of maximally entangled states generation. The system discussed here could be a part of a more complicated model that allows 
generation of maximally entangled states with the help of the above mentioned purification method.

\section{Conclusions}
We have discussed the model of a coherently driven NC with a nonlinear
internal coupling showing that for sufficiently small values of the
coupling constants we can apply the \textit{nonlinear scissors}
method. Consequently, the system's evolution becomes closed inside
a set of three states. Moreover, which seems to be the most interesting,
we have shown that our system can be treated as a \textit{maximally
entangled states} generator, in particular, it can generate the Bell-like states. 
We have also discussed the fragility of the system
to the damping processes. The model proposed here, due to the nonlinear
character of the internal coupling, seems to be more suitable for
practical realization than that involving only linear interactions. As
we discuss the context of practical feasibility of our model we should
keep in mind the assumptions of very weak interactions. Consequently,
the model could be practically realizable if we assume sufficiently
large values of the nonlinearities discussed here. This fact indicates
that the most suitable for practical realization would be systems
based on the \textit{electromagnetically induced transparency}
phenomenon or \textit{atomic dark resonances}. These processes have
been proposed by Schmidt and Imamo\u{g}lu \cite{SI96,ISWD97,ISWD98}
and experimentally observed \cite{KZ03}. For models involving
these effects one could expect the possibility of producing the so-called
\textit{giant Kerr nonlinearities}. Imamo\u{g}lu has estimated
that for such systems we can achieve $\chi\sim 10^8$ rad/sec
\cite{ISWD97,ISWD98}. So, we see that the system discussed in this
paper can be treated at least as a potential source of the Bell-like
states. Moreover, it could be implemented as a part of a more
complicated model interesting from the quantum information theory
point of view, keeping in mind all obstacles and limitations concerning its practical realization.

\begin{acknowledgments}
The authors wish thank Prof. R.Tana\'s and Prof. A. Miranowicz for their valuable and inspiring discussions and suggestions.

This work was supported by the Polish State Committee for
Scientific Research under grant No. 1 P03B 064 28. 
\end{acknowledgments}

\end{document}